\documentclass[12pt]{iopart}
\usepackage{epsfig,graphicx,tabularx}
\usepackage{psfrag}
\usepackage{graphicx}
\usepackage{graphics}
\usepackage{epsfig}
\usepackage{bm}
\usepackage{verbatim,color,ulem}

\newcommand{\beq}{\begin{equation}}
\newcommand{\eeq}{\end{equation}}
\newcommand{\beqa}{\begin{eqnarray}}
\newcommand{\eeqa}{\end{eqnarray}}

\newcommand{\be}{\begin{equation}}
\newcommand{\ee}{\end{equation}}
\newcommand{\bea}{\begin{eqnarray}}
\newcommand{\eea}{\end{eqnarray}}

\begin{document}

\title{Condensation and superfluidity of dilute Bose gases with 
finite-range interaction}

\author{A. Tononi}
\address{Dipartimento di Fisica e Astronomia ``Galileo Galilei'' 
and CNISM, Universit\`a di Padova, Via Marzolo 8, I-35131 Padova, Italy}
\author{A. Cappellaro}
\address{Dipartimento di Fisica e Astronomia ``Galileo Galilei'' 
and CNISM, Universit\`a di Padova, Via Marzolo 8, I-35131 Padova, Italy}
\author{L. Salasnich}
\address{Dipartimento di Fisica e Astronomia ``Galileo Galilei'' 
and CNISM, Universit\`a di Padova, Via Marzolo 8, I-35131 Padova, Italy}
\address{Istituto Nazionale di Ottica (INO) del Consiglio Nazionale 
delle Ricerche (CNR), Via Nello Carrara 1, I-50019 Sesto Fiorentino, Italy}

\date{\today}

\begin{abstract}
We investigate an ultracold and dilute Bose gas by taking into account 
a finite-range two-body interaction. The coupling constants of the resulting 
Lagrangian density are related to measurable scattering parameters by 
following the effective-field-theory approach. A perturbative scheme is 
then developed up to the Gaussian level, where both quantum and thermal 
fluctuations are crucially affected by finite-range corrections.
In particular, the relation between spontaneous symmetry breaking and the 
onset of superfluidity is emphasized by recovering the renowned 
Landau's equation for the superfluid density in terms of the condensate one. 
\end{abstract}


\maketitle
\section{Introduction}

Since the seminal investigation on liquid Helium by H. Kamerlingh Onnes, 
the research on low-temperature physics has been focused on the understanding 
and engineering of superfluid states of matter \cite{tilley}. 
Nowadays, superfluids and superconducting materials are at the core
of a new technological revolution centered around the development 
of quantum devices \cite{pezze2018,streltsov2017,kimble2018}. 

From a theoretical point of view, an enormous effort has been devoted to 
finally provide a microscopic theory accounting for the transition 
between a normal state and a superfluid one, where dissipationless 
flow occurs. This research oscillated between  the necessity of statistical 
mechanics to identify general mechanisms leading to supertransport and 
the interests of condensed matter theorists and material scientists on 
peculiar setups with unique properties. Concerning metallic 
superconductors, the theoretical investigation reached one of its 
peaks with the BCS theory and its consequent 
refinements \cite{bcs,varlamov-book}. 

Moving to the superfluid side, liquid Helium remains for decades 
the only efficient platform to probe the markers of superfluid behaviour, 
like the absence of viscous forces and the vorticity 
quantization \cite{hohenberg1965,anderson1966}. 
The formulation of a microscopic theory for the superfluid 
phase of liquid Helium proves to be an exceptionally demanding task. 
Up to now, a reliable picture can be achieved via \textit{ab initio} 
numerical simulations such as Path-Integral Monte Carlo algorithms
\cite{ceperley1995}. Indeed, liquid Helium has to be classified as a 
strongly-correlated system, due to the experimental values at play 
for density and interaction strength. In this situation, 
even identifying a \textit{smallness} parameter is non trivial, 
preventing the effective implementation of a perturbative 
expansion \cite{schmitt-book}.

The two-fluid phenomenological approach by 
Landau \cite{landau1941,landau1981} has been much more fruitful, 
enabling the derivation of a self-consistent hydrodynamic 
theory over few crucial assumptions. Remarkably, the original formulation 
of Landau did not rely upon an atomic point of view and, 
moreover, did not invoke any symmetry breaking related to the modern 
characterization of phase transitions \cite{leggett-book}.

Since the pioneering guess of London in 1938 
\cite{london1938-1,london1938-2}, it has been a common approach to 
interpret the superfluid transition in terms of Bose-Einstein
condensation, where a macroscopic fraction of Helium atoms can be described 
by a macroscopic wavefunction. Unfortunately, 
analytical result can be obtained only in the weakly interacting limit, 
which does not hold up to the experimental values for Helium.
Applied to Helium, the resulting picture is only qualitative since, 
for instance, it does not
even manage to capture the peculiar rotonic minimum of the excitation spectrum.

In 1995, the experimental realization of a Bose-Einstein condensates 
\cite{leggett-book} changed the scenario in a crucial way: 
for the first time, the predictions of the Bogoliubov theory 
\cite{bogoliubov1947} have been checked in actual weakly-interacting 
quantum gases. At the same time, within a field-theory approach, 
it is possible to recover the Landau main results moving from a 
microscopic Lagrangian for ultracold atoms. 

The several successful theoretical studies based on the Bogoliubov 
framework move from the crucial assumption that the  \textit{true} 
atom-atom interaction can be replaced by a contact (i.e. zero-range) 
pseudopotential whose strength is given by the s-wave scattering length $a_s$
\cite{huang-book,stringari-book}. 
The resulting thermodynamics is universal since there is no dependence 
on the potential shape, with only $a_s$ playing a relevant role. 
The same point can be made for transport quantities as 
the superfluid fraction. Despite the many achievements of this strategy, 
current experiments deal with higher density setups, reduced 
dimensionalities and more complex interactions 
\cite{hadzibabic2009,dalibard2011}.
Thus, it is pressing to extend the usual two-body zero-range framework
in order to capture more realistic and interesting experimental regimes. 
\begin{figure}[ht!]
\centering
\includegraphics[scale=0.50]{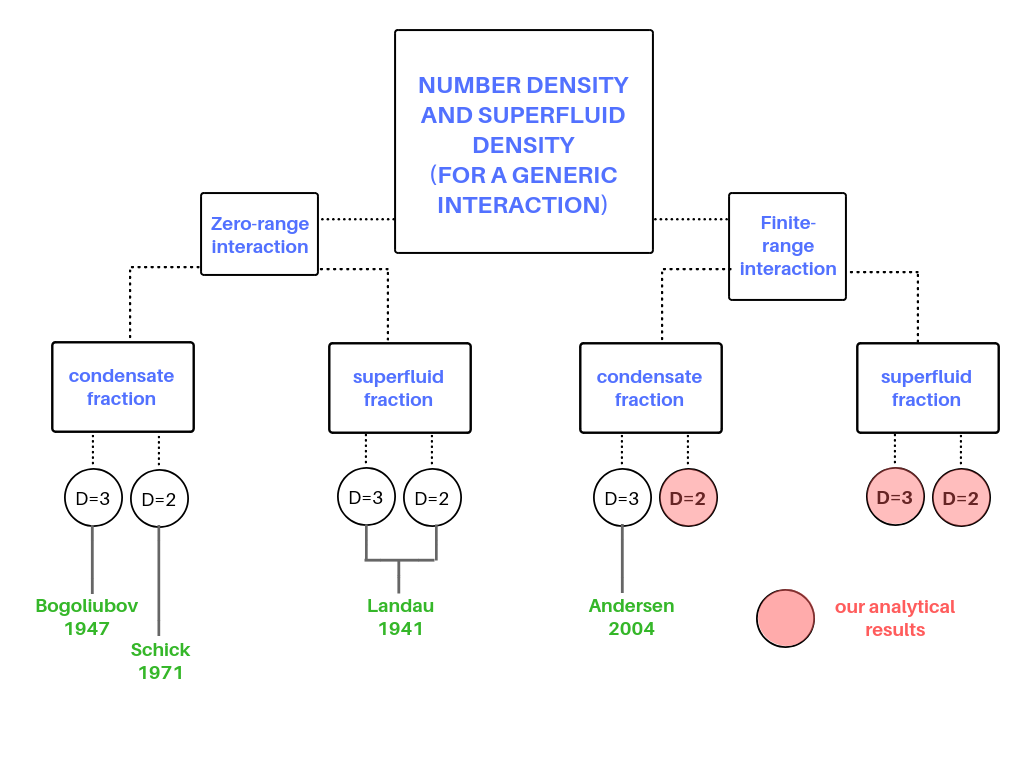}
\caption{Flowchart diagram showing some the key theoretical developments related
to our analysis of three- and two-dimensional Bose gases with a finite-range
interaction potential.}
\label{fig:1}
\end{figure}
Within a functional integration formalism, atoms are represented by a 
bosonic field whose dynamics is governed by a microscopic 
interacting Lagrangian density. The coupling constants of the 
finite-range theory can be determined in terms of the s-wave scattering 
parameters, namely $a_s$ and the corresponding effective range $r_e$. In 
\cite{braaten2001,andersen2004,cappellarosalasnich2017,yukalov},
the finite-range thermodynamics is derived up to the Gaussian level for 
a three dimensional uniform Bose gas, while the non-trivial 
case of two spatial dimensions is addressed in \cite{salasnich2017bis,bean}. 
In Fig. \ref{fig:1} we report a visual summary of the major 
analytical approaches to modelling bosonic quantum gases. 

A similar analysis concerning the superfluid properties of a finite-range 
theory is still missing and it is the main subject of this work. 
By adopting a functional integration point of view as in
\cite{cappellarosalasnich2017,salasnich2017bis}, we are going to  
show that both condensate and superfluid depletion are modified 
by the finite-range character of the two-body interaction. 
Moreover, they are not independent from each other but the
familiar Landau equation for the superfluid density can be formulated in terms
of the condensate one.

The paper is organized as follows: in the next section we discuss 
the Landau two-fluid model from a field-theory perspective moving 
from a microscopic Langrangian density. 
The key point consists in properly accounting the different response 
of normal and superfluid components to a Galilean boost \cite{schakel}. 
Technically, this corresponds to performing a phase twist on 
the order parameter \cite{fisher,taylor2006}, which has to be considered 
throughout the whole perturbative expansion. By following this scheme, 
we will derive a Landau-like equation relating superfluid 
and condensate density. Our calculation are carried on in a generic 
dimension $D$, simplifying the application of dimensional regularization 
to heal the UV-divergent zero-point energy. 
We then specify to the cases $D=3$ and $D=2$.

\section{Two-fluid model of a superfluid}

The thermodynamic properties of a physical system can be described 
calculating the grand canonical partition function ${ \cal Z}$, which is related 
to the grand potential $\Omega$ by
\begin{equation}
\label{grandpotential}
\Omega=-\frac{1}{\beta} \ln({ \cal Z }).
\end{equation}
We consider a uniform $D$-dimensional Bose gas of identical cold atoms 
described by the complex scalar field $\psi(\vec{r},\tau)$. We calculate 
the grand canonical partition function ${ \cal Z}$ as the functional integral 
\begin{equation}
\label{partfunction}
{ \cal Z }=\int{{ \cal D} [\bar{\psi},\psi] \; e^{-\frac{S[\bar{\psi},\psi]}{\hbar}}},
\end{equation}
where 
\begin{equation}
\label{action}
S[\bar{\psi},\psi] = \int_{0}^{\beta\hbar} d\tau \int_{L^D} d^D r \; 
{ \cal L }(\bar{\psi},\psi) 
\end{equation}
is the Euclidean action and $\beta = 1/(k_{B}T)$, with $k_{B}$ the 
Boltzmann constant and $T$ the absolute temperature. 
We introduce the non-relativistic Lagrangian
\beqa
\label{lagrangian}
{ \cal L } =& \bar{\psi}(\vec{r},\tau) \bigg(\hbar\partial_{\tau}-
\frac{\hbar^2 \nabla^2}{2m}-\mu\bigg) \psi(\vec{r},\tau) \nonumber \\
&+ \frac{1}{2}\int d^{D}r \, ' \ |\psi(\vec{r} \, ',\tau)|^{2} \ 
V(|\vec{r}-\vec{r }\, '|) \ |\psi(\vec{r},\tau)|^{2},
\eeqa
where $\mu$ is the chemical potential and $V(|\vec{r}-\vec{r }\, '|)$ is 
the generic interaction between bosons, assuming it is dependent 
only on the relative distance $|\vec{r}-\vec{r }\, '|$.  

In the context of the two-fluid model of a superfluid 
\cite{landau1941,landau1981} we describe the fluid behavior of the system 
as composed by a mixture of a normal component and a superfluid component. 
In particular, we consider the normal part as a fluid current 
moving with velocity $\vec{v}$ with respect to the laboratory frame of 
reference. Working with imaginary time, we describe this motion by 
substituting the time derivative in the Lagrangian (\ref{lagrangian}) 
with the Lagrangian fluid derivative
\begin{equation}
\label{eulerianderivative}
{\partial_{\tau}} \rightarrow {\partial_{\tau}} - i \vec{v} \cdot \vec{\nabla}
\end{equation}
Moreover, since the superfluid part does not exchange momentum with the 
normal part, we impose a superfluid current with a phase twist of the bosonic field \cite{fisher}
\begin{equation}
\label{fieldtrasnformation}
\psi(\vec{r},\tau) \rightarrow e^{i \frac{m\vec{v}_{s}\cdot\vec{r}}{\hbar} } 
\psi(\vec{r},\tau)
\end{equation}
Substituting these expressions in the Lagrangian (\ref{lagrangian}) we obtain
\beqa
{ \cal L } = \bar{\psi}(\vec{r},\tau) \bigg( \hbar \partial_{\tau} 
- \frac{\hbar^2 \nabla^2}{2m} -\mu_{e}  + (\vec{v}-\vec{v}_{s})
\cdot(-i\hbar\vec{\nabla}) \bigg) \psi(\vec{r},\tau) \nonumber 
\\ 
+ \frac{1}{2}\int d^{D}r \, ' \ |\psi(\vec{r} \, ',\tau)|^{2} \ 
V(|\vec{r}-\vec{r }\, '|) \ |\psi(\vec{r},\tau)|^{2}, 
\eeqa
where we define the effective chemical potential $\mu_{e}$ as \cite{schakel}
\begin{equation}
\mu_{e} = \mu - \frac{1}{2}m \vec{v}_{s} \cdot (\vec{v}_{s} - 2\vec{v})
\end{equation}
Considering that in the condensed phase the $\mbox{U}(1)$ global symmetry 
is spontaneously broken we use the bosonic field parametrization
\begin{equation}
\label{parametrizationveta}
\psi(\vec{r},\tau)=\psi_{0} +\eta(\vec{r},\tau)
\end{equation}
where $\eta(\vec{r},\tau)$ is the complex field describing the 
fluctuation around the uniform field configuration $\psi_{0}$, which 
represents the order parameter of the condensate phase transition. 
Substituting the parametrization (\ref{parametrizationveta}) in the 
action (\ref{action}) and keeping only quadratic terms in the fluctuation 
field $\eta(\vec{r},\tau)$ we obtain the homogeneous system action
\beq
S_{0}= \beta\hbar L^D(-\mu_{e} \psi_{0}^2 +\frac{1}{2}g_{0} \psi_{0}^4)
\eeq
and we calculate the Gaussian action in the Fourier space as \cite{nagaosa}
\beq 
S_{g}[\bar{\eta},\eta] =  \frac{\hbar}{2}\sum_{k}  \left( \bar{\eta}(k) 
\   \eta(-k) \right) 
\label{actionGaussian}
\ {\cal M} \ 
\left( \begin{array}{c} 
\eta(k) \\
\bar{\eta}(-k)
\end{array} \right) 
\eeq
where $k = (\vec{k},\omega_{n})$ are the $D+1$ wavevectors, $\omega_{n} = \frac{2 \pi n}{\beta \hbar}$ are the bosonic Matsubara frequencies and
\beq 
{\cal M} = 
\left( \begin{array}{c c } 
{\cal M }_{11} 
& {\cal M }_{12}  \\ 
{\cal M }_{21}  & {\cal M }_{22}  
\\ 
\end{array} \right)  
\eeq
with
\beqa
{\cal M }_{11} &=-i\hbar\omega_{n}+\frac{\hbar^2k^2}{2m}-\mu_{e} + g_{0} \psi_{0}^2+ \psi_{0}^2 \, { \tilde V}(\vec{k}) + \hbar(\vec{v}-\vec{v}_{s}) \cdot \vec{k} \nonumber \\
{ \cal M }_{22} &= +i\hbar\omega_{n}+\frac{\hbar^2k^2}{2m}-\mu_{e} + g_{0} \psi_{0}^2+ \psi_{0}^2 \, { \tilde V}(\vec{k}) - \hbar(\vec{v}-\vec{v}_{s}) \cdot \vec{k} \nonumber \\
{\cal M }_{12} &= {\cal M }_{21} = \psi_{0}^2 { \tilde V}(\vec{k})
\eeqa
Performing the functional integration of the Gaussian action (\ref{actionGaussian}) we obtain the partition function (\ref{partfunction}) and the grand potential (\ref{grandpotential}). In particular, we find that the grand potential $\Omega$ is the sum of two contributions
\beq
\label{grandpotential0g}
\Omega = \Omega_{0} + \Omega_{g}
\eeq
where
\beq
\label{Omega0}
\Omega_{0} = L^D(-\mu_{e} \psi_{0}^2 +\frac{1}{2}g_{0} \psi_{0}^4)
\eeq
is the order parameter grand potential and
\begin{equation}
\label{gaussiangp}
\Omega_{g} =  \frac{1}{2 \beta} \sum_{\vec{k},\omega_{n}} \ln[\beta^2(E_{\vec{k}}(\psi_{0}^{2})^2 - (i\hbar\omega_{n}-\hbar(\vec{v}-\vec{v}_{s}) \cdot \vec{k} \ )^{2})]
\end{equation}
is the Gaussian contribution to the grand potential, where
\beq
\label{excitationspectrum}
E_{\vec{k}}(\psi_{0}^{2}) =\sqrt[]{\left( \frac{\hbar^2k^2}{2m}-
\mu_{e} + g_{0} \psi_{0}^2+ \psi_{0}^2 \, { \tilde V}(\vec{k}) 
\right)^{2}-\big(\psi_{0}^2 \, { \tilde V}(\vec{k})\big)^{2}}
\eeq
is the excitation spectrum of the Bose gas. 
{The Hugenholtz-Pines theorem 
is guaranteed imposing the saddle point condition 
$\partial \Omega / \partial \psi_{0}= 0$, for which the excitation 
spectrum becomes gapless and the chemical potential reads
\begin{equation} \label{saddlepoint}
\mu_{e}= g_{0} \psi_{0}^{2}
\end{equation} 
Moreover, we identify the condensate density as $n_{0}=\psi_{0}^2$. 
Notice that, within the Gaussian approximation of the action, 
the excitation spectrum does not contain the anomalous density, 
which is instead included by adopting the next-next-to-leading 
Hartree-Fock-Bogoliubov scheme \cite{boudjemaa2011,boudjemaa2012,boudjemaa2017}.} 
The initial assumption that the real space interaction depends only 
on the distance between bosons implies that the interaction potential 
is left unchanged by a reflection of the momenta: 
${ \tilde V}(\vec{k})={ \tilde V}(-\vec{k})$.
Due to this property we are able to perform the summation 
over Matsubara frequencies $\omega_{n}$ in the Gaussian grand potential 
(\ref{gaussiangp}) \cite{schakel,altland}, obtaining the grand potential 
$\Omega$ in the form
\beq
\label{grandpotential0g0gT}
\Omega = \Omega_{0} + \Omega_{g}^{(0)} + \Omega_{g}^{(T)}
\eeq
where $\Omega_{0}$ is given by Eq. (\ref{Omega0}),
\beq
\label{Omegag0}
\Omega_{g}^{(0)} = \frac{1}{2} \sum_{\vec{k}} E_{\vec{k}}(\psi_{0}^{2})
\eeq
is the zero-temperature Gaussian contribution, 
written as the sum of noninteracting elementary excitations 
with spectrum $E_{\vec{k}}(\psi_{0}^{2})$, and 
\beq
\label{OmegagT}
\Omega_{g}^{(T)} =  \frac{1}{\beta} \sum_{\vec{k}} \ln(1-e^{-\beta( 
E_{\vec{k}}(\psi_{0}^{2}) + \hbar(\vec{v}-\vec{v}_{s}) \cdot \vec{k})})
\eeq
is the finite-temperature Gaussian contribution. 
{The beyond-mean-field Gaussian equation of state with 
finite-range interaction has been analyzed in previous papers 
\cite{braaten2001,cappellarosalasnich2017,salasnich2017bis}. 
For the sake of completeness we report the main results in the Appendix.}

We now calculate the superfluid density $n_{s}$ of the system with a 
self-consistent approach, employing the Gaussian grand potential 
(\ref{grandpotential0g0gT}). In particular, we will identify $n_{s}$ from 
the calculation of the total momentum density $\vec{\cal P }$ of the fluid, 
which is obtained as follows
\begin{equation}
\vec{\cal P } = \frac{1}{L^{D}} \frac{\partial\Omega(\mu_{e},\psi_{0})}
{\partial (-\vec{v})} \bigg |_{ {\mu_{e} =g_{0} \psi_{0}^2 = g_{0} n_{0} } }
\end{equation}
where we take the derivative of the grand potential with respect 
to the velocity $- \vec{v}$ first and then we substitute the mean field 
value of the chemical potential $\mu_{e} =g_{0}\psi_{0}^2$. We find that the 
momentum density $\vec{\cal P }$ is given by
\begin{equation}
\label{momentumdensity}
\vec{\cal P } =  \vec{\cal P }_{0} + \vec{\cal P }_{g}^{(0)} + 
\vec{\cal P }_{g}^{(T)}
\end{equation}
where, since the grand potential (\ref{grandpotential0g0gT}) is given 
by the sum of three sum contributions, the three terms of the momentum 
density $\vec{\cal P }$ are defined accordingly. In particular, we find
\beq
\label{P0}
\vec{\cal P }_{0} = n_{0} m \vec{v}_{s}
\eeq
The Gaussian zero-temperature contribution $\vec{\cal P }_{g}^{(0)}$ is
\beq
\label{Pg0}
\vec{\cal P }_{g}^{(0)} = f_{g}^{(0)}(n_{0}) \; m \vec{v}_{s}
\eeq
where we define the zero-temperature number density contribution 
$f_{g}^{(0)}(n_{0})$ as
\beq
\label{fg0}
f_{g}^{(0)} (n_{0})= \frac{1}{2 L^{D}} \sum_{\vec{k}}  \frac{1}{E_{\vec{k}}
(n_{0})} \bigg(\frac{\hbar^2k^2}{2m}+ n_{0} {\tilde V}(\vec{k}) \bigg)
\eeq
The Gaussian thermal contribution $\vec{\cal P }_{g}^{(T)}$ to the momentum 
density (\ref{momentumdensity}) is more involved:
\begin{equation}
\label{PgT}
\vec{\cal P }_{g}^{(T)} = \frac{1}{L^{D}} \sum_{\vec{k}} 
\frac{1}{e^{\beta( E_{\vec{k}}(\psi_{0}^2) + \hbar(\vec{v}-\vec{v}_{s}) \cdot \vec{k})}-1} 
\bigg[  \frac{\partial E_{\vec{k}}(\psi_{0}^2)}{\partial \mu_{e}}  \ 
\frac{\partial\mu_{e}}{\partial (-\vec{v})}  - \hbar \vec{k}  \bigg] 
\bigg |_{ { \mu_{e} =g_{0} \psi_{0}^2 = g_{0} n_{0}  } }
\end{equation}
Assuming that the difference between the velocity $\vec{v}$ of the normal fluid and the velocity $\vec{v}_{s}$ of the superfluid is small, we can expand the exponential and, taking into account that some terms are zero for symmetry reasons, we obtain
\begin{equation}
\label{PgT2}
\vec{\cal P }_{g}^{(T)} = f_{g}^{(T)}(n_{0}) \; m \vec{v}_{s} + n_{n}(n_{0},T) \; m (\vec{v}-\vec{v}_{s})
\end{equation}
where $f_{g}^{(T)}(n_{0})$ is the Gaussian thermal density contribution
\beq
\label{fgT}
f_{g}^{(T)} (n_{0}) = \frac{1}{L^{D}} \sum_{\vec{k}}  \frac{1}{E_{\vec{k}}(n_{0})} \bigg(\frac{\hbar^2k^2}{2m}+ n_{0} {\tilde V}(\vec{k}) \bigg) \frac{1}{e^{\beta E_{\vec{k}}(n_{0})}-1} 
\eeq
and we have defined the normal density of the fluid as
\begin{equation}
\label{schakelnormaldensity}
n_{n}(n_{0},T) = \frac{\beta \hbar^2}{m D L^D} \sum_{\vec{k}}  k^{2} \ \frac{e^{\beta E_{\vec{k}}(n_{0})}}{(e^{\beta E_{\vec{k}}(n_{0})}-1)^{2}} 
\end{equation}
Notice that, in the thermodynamic limit $L^D \rightarrow \infty$, the normal fluid density (\ref{schakelnormaldensity}) is fully consistent with the familiar Landau result \cite{landau1981}. In conclusion, putting together the contributions (\ref{P0}), (\ref{Pg0}) and (\ref{PgT2}) we rewrite the momentum density $\vec{\cal P }$ as 
\begin{equation}
\vec{\cal P } = [n_{0} + f_{g}^{(0)}(n_{0}) + f_{g}^{(T)}(n_{0})] m \vec{v}_{s} + n_{n}(n_{0},T) \; m (\vec{v}-\vec{v}_{s})
\end{equation}
In the square bracket we identify the number density $n(n_{0}, T)$, expressed as a function of the condensate number density $n_{0}$ and the temperature $T$ as follows \cite{schakel}
\beqa
\label{number density}
n(n_{0}, T) = - \frac{1}{L^{D}} \frac{\partial\Omega(\mu_{e},\psi_{0},T)}{\partial \mu_{e}} \bigg|_{ \mu_{e} =g_{0}n_{0}  } = n_{0} + f_{g}^{(0)} (n_{0}) + f_{g}^{(T)} (n_{0})
\eeqa
We remark that we take the derivative of the grand potential with respect to the chemical potential first and then we substitute the mean field value of the chemical potential $\mu_{e} =g_{0}\psi_{0}^2$, with the identification for the condensate density $n_{0}=\psi_{0}^2$. This procedure can be justified considering that the same procedure is implemented to calculate the condensate fraction of a noninteracting Bose gas \cite{salasnichtoigo}. 
With this identification, we express the momentum density $\vec{\cal P }$ as
\begin{equation}
\vec{\cal P } = n(n_{0}, T) m \vec{v}_{s} + n_{n}(n_{0},T) \; m (\vec{v}-\vec{v}_{s})
\end{equation}
Finally, we identify the superfluid density $n_{s}$ as
\begin{equation}
\label{superfluid density}
n_{s} = n(n_{0}, T) - n_{n}(n_{0}, T)
\end{equation}
which allows us to express the momentum density $\vec{\cal P }$ as the sum of the momentum density $n_{s} m \vec{v}_{s}$ of the superfluid part of the fluid and the momentum density $n_{n} \left( n_{0},T \right) m \vec{v}$ of the normal part of the fluid, namely
\beq
\vec{\cal P } = n_{s} m \vec{v}_{s} +  n_{n} \left( n_{0},T \right) m \vec{v}
\eeq
We emphasize that Eq. (\ref{superfluid density}) constitutes the main result of this paper, since it highlights the non-trivial relationship between the superfluid density $n_{s}$, the condensate density $n_{0}$ and the temperature $T$. This result may be regarded as the explicit formulation, at a Gaussian level, of the Josephson relation \cite{josephson1966}.

Notice that our number density (\ref{number density}) and the superfluid fraction (\ref{superfluid density}) are equivalent to the result obtained with Beliaev diagrammatic technique in reference \cite{svistunov} if we approximate $n_{0} \approx n$ in Eqs. (\ref{fg0}), (\ref{fgT}) and (\ref{schakelnormaldensity}) and we consider the zero-range interaction ${\tilde V}(\vec{k}) = g_{0}$.
In the following section we implement the superfluid density calculation for bosons with finite-range interaction in three- and two- dimensional systems.

\section{Superfluid density of bosons with finite-range 
interaction}

In order to obtain explicit formulas for the superfluid density $n_{s}$, 
in this section we shall implement Eq. (\ref{superfluid density}) in the 
three- and in the two- dimensional Bose gas, considering the explicit form 
of the interaction $V(\vec{r})$. The usual approximation to study a 
weakly-interacting Bose gas of ultracold atoms is constituted by the 
zero-range interaction $V(\vec{r}) = g_{0} \, \delta(\vec{r})$, which in the 
Fourier space gives
\begin{equation}
\label{contact interaction}
    {\tilde V}(\vec{k}) = g_{0}
\end{equation}
Here we improve this approximation, considering the finite-range effective 
interaction
\begin{equation}
\label{finite-range interaction}
    {\tilde V}(\vec{k}) = g_{0} + g_{2} k^2
\end{equation}
which is obtained adding to the zero-range interaction strength $g_{0}$ 
the first nonzero correction in the gradient expansion of the real interaction 
potential $V(|\vec{r}-\vec{r} \, '|)$, namely $g_{2}k^2$, where we define
\begin{equation}
g_{0} = \int d^{D}r \ V(r), \qquad g_{2} = - \frac{1}{2} \int d^{D}r \ r^{2} \ V(r)
\end{equation}
In the three dimensional case, the values of the couplings $g_{0}$ and 
$g_{2}$ are determined with the scattering theory in terms of the s-wave scattering length $a_{s}$ and the effective range $r_{e}$ as follows \cite{cappellarosalasnich2017}
\begin{equation}
\label{g23D}
g_{0} = \frac{4 \pi \hbar^2 a_{s}}{m},  \qquad  g_{2}= \frac{2 \pi \hbar^2 a_{s}^{2} r_{e}}{m}
\end{equation}
In the two-dimensional case we choose the zero-range interaction coupling $g_{0}$ according to reference \cite{astrakharchik2009} and we derive the coupling $g_{2}$ from the definition of the characteristic range $R = 2 \, \sqrt{|g_{2}/g_{0}|}$ discussed in \cite{salasnich2017bis}, obtaining
\begin{equation}
\label{g0g22D}
g_{0} =  \frac{4 \pi \hbar^{2}}{m} \frac{1}{|\ln(n a_{s}^{2})|} \qquad g_{2} = \frac{\pi \hbar^{2}}{m} \frac{R^{2}}{|\ln(n a_{s}^{2})|}
\end{equation}

We now explicitly implement the superfluid density $n_{s}$ calculation for the finite-range effective interaction (\ref{finite-range interaction}). According to Eq. (\ref{superfluid density}), $n_{s}$ is obtained by subtracting the normal density $n_{n}(n_{0},T)$ from the number density $n(n_{0},T)$. We first calculate the number density $n(n_{0},T)$, which is given as the sum of the three contributions of Eq. (\ref{number density}). The first contribution is the condensate density $n_{0}$. The second contribution is the zero-temperature Gaussian contribution to the normal density, namely
\begin{equation}
\label{fg03dfiniterangetoregularize}
f_{g}^{(0)} (n_{0})= \frac{S_{D}}{2 (2 \pi)^{D}} \int_{0}^{+\infty} dk \, k^{D-1}    \frac{1}{E_{k}(n_{0})} \bigg( \frac{\hbar^2k^2}{2m} \tilde{\lambda}(g_{2},n_{0}) + g_{0} n_{0} \bigg) 
\end{equation}
where $S_{D} = {2 \pi^{D/2}}/{\Gamma[D/2]}$ is the $D$-dimensional solid angle, in which $\Gamma[x]$ is the Euler gamma function and where we define the excitation spectrum
\begin{equation}
\label{excitationspectrumfiniterange}
E_{k}(n_{0}) = \sqrt[]{ \frac{\hbar^2k^2}{2m}\bigg(\frac{\hbar^2k^2}{2m} \lambda(g_{2},n_{0}) + 2 n_{0} g_{0} \bigg) }
\end{equation}
with
\beqa
\label{lambda}
\lambda(g_{2},n_{0}) = 1+ \frac{4m}{\hbar^2} g_{2} n_{0} \qquad  \tilde{\lambda}(g_{2},n_{0}) = 1 + \frac{2m}{\hbar^2} g_{2} n_{0}
\eeqa
Notice that the excitation spectrum $E_{k}(n_{0})$ reproduces the familiar Bogoliubov spectrum if the zero-range interaction is restored by putting $g_{2}=0$.
Since $f_{g}^{(0)} (n_{0})$ is ultraviolet divergent, we will regularize it using dimensional regularization \cite{thooft1972,leibbrandt1975}. In particular, we obtain an adimensional integral using the integration variable $t= \hbar^2k^2 \lambda(g_{2},n_{0}) / (4 m g_{0} n_{0}) $, then we extend the spatial dimension $D$ to the complex value ${\cal D} = D - \varepsilon$.  We remark that this additional step is needed because the dimensional regularization procedure is not always able to heal the ultraviolet divergence of the integrals \cite{salasnichtoigo}. After the integration, we obtain $f_{g}^{(0)} (n_{0})$ in the form
\beqa
\label{fcondfracfiniterange} \nonumber
f_{g}^{(0)} (n_{0})= \frac{\kappa^{\varepsilon} \lambda(g_{2},n_{0})^{1/2} }{4 \Gamma({\cal D}/2)} \  \bigg( \frac{m g_{0} n_{0}}{ \pi \hbar^2 \lambda(g_{2},n_{0})} \bigg)^{{\cal D}/2} \ \bigg[2   \frac{\tilde{\lambda}(g_{2},n_{0})}{\lambda(g_{2},n_{0})} \cdot \\ B\bigg(\frac{{\cal D}+1}{2},\frac{-{\cal D}}{2} \bigg) + B \bigg(\frac{{\cal D}-1}{2},\frac{2-{\cal D}}{2} \bigg) \bigg]
\eeqa
where $B(x,y)$ is the Euler Beta function and $\kappa$ is an ultraviolet cutoff wavevector introduced for dimensional reasons.

The third term in the number density of Eq. (\ref{number density}) is  the finite-temperature Gaussian contribution $f_{g}^{(T)} (n_{0})$, which, unlike the zero-temperature one, is convergent. However, this integral can be calculated analitically only in the low-temperature regime, where it is useful to introduce the integration variable $x = \beta E_{k}(n_{0})$. Doing so in Eq. (\ref{fgT}) in which the finite-range interaction (\ref{finite-range interaction}) is substituted, we get
\begin{equation}
\label{fgtcondfracfiniterange}
f_{g}^{(T)} (n_{0}) = \frac{S_{D}}{(2 \pi)^{D} k_{B} T} \int_{0}^{+\infty} dx \ \frac{dk (x)}{dx} \frac{k(x)^{D-1} }{x ( e^{x} - 1)} \bigg( \frac{\hbar^2 k(x)^2}{2m} \lambda(g_{2},n_{0}) + n_{0} g_{0} \bigg)
\end{equation} 
with
\begin{equation}
\label{kfiniterange}
k(x) = \sqrt[]{\frac{2m n_{0} g_{0} } {\hbar^2 \lambda(g_{2},n_{0})} } \ \sqrt[]{-1 + \sqrt[]{1+ \frac{(k_{B} T)^2 \lambda(g_{2},n_{0}) x ^{2}}{n_{0}^2 g_{0}^2 }}}
\end{equation}
Before substituting the spatial dimension $D$ to calculate 
explicitly the number density contributions obtained in the previous 
equations, let us also calculate the normal density $n_{n}(n_{0},T)$, 
which is given by Eq. (\ref{schakelnormaldensity}) where the finite-range interaction (\ref{finite-range interaction}) is substituted. In analogy with the finite-temperature density contribution $f_{g}^{(T)} (n_{0})$, the normal density $n_{n}\left( n_{0},T \right)$ can be calculated analytically only in the low-temperature regime: as before we introduce the integration variable $x = \beta E_{k}(n_{0})$, obtaining
\begin{equation}
\label{normaldensityfiniterange}
n_{n}\left( n_{0},T \right) = \frac{\beta \hbar^{2} S_{D}}{m D (2 \pi)^{D}} \int_{0}^{+\infty}d x\ \frac{d k}{dx}(x) \ k(x)^{D+1} \ \frac{e^{x}}{(e^{x}-1)^{2}} 
\end{equation}
where $k(x)$ is given again by Eq. (\ref{kfiniterange}). We now calculate explicitly the number density $n(n_{0},T)$ and the superfluid density $n_{s}(n_{0},T)$ in $D=3$ and in $D=2$

\subsection{$D=3$}

\begin{figure}[ht!]
\centering
\includegraphics[scale=0.80]{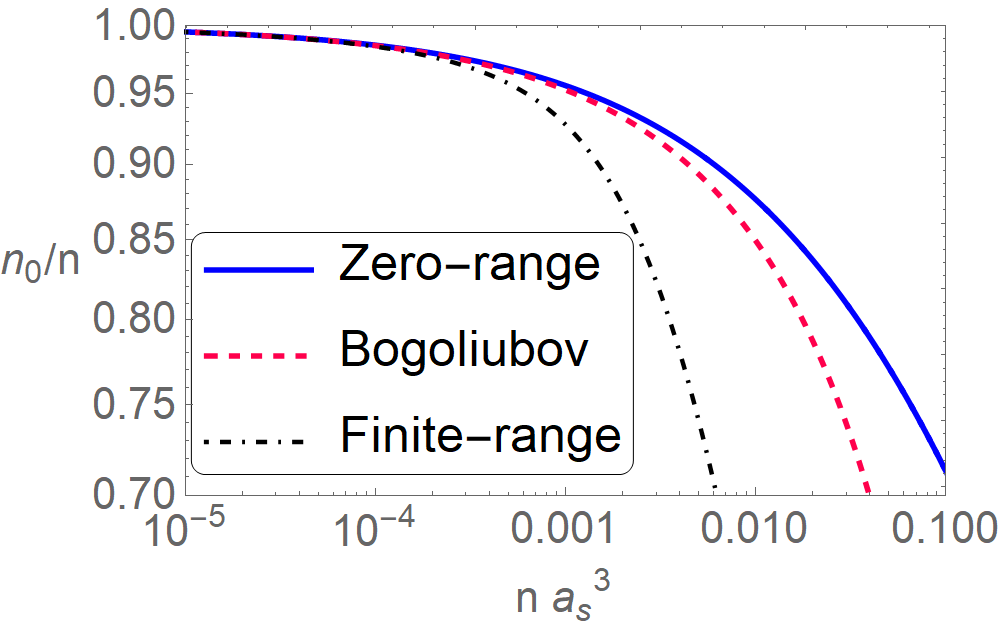}
\caption{Zero-temperature condensate fraction $n_{0}/n$ as a function of the 
gas parameter $n a_{s}^{3}$ in $D=3$. Here we represent the finite-range 
condensate fraction (black dot-dashed line) for the effective range value 
$r_{e}/a_{s} = -10$ in comparison with the result for zero-range interaction 
(blue solid line) and the Bogoliubov's result (red dashed line). 
Notice that, with a weak dependence on the choice of $r_{e}$, the finite-range 
correction becomes relevant for values of the gas parameter $n a_{s}^{3}$ 
greater than $10^{-3}$.}
\label{fig:2}
\end{figure}

As a preliminar result for the superfluid density $n_{s}$, we employ 
Eq. (\ref{number density}) to calculate the density $n \left( n_{0},T \right)$ 
of bosons with finite-range interaction. In $D=3$, the zero-temperature 
Gaussian contribution  $f_{g}^{(0)} (n_{0})$ is regularized simply by the means 
of dimensional regularization, therefore we put $\varepsilon = 0$ into the 
equation (\ref{fcondfracfiniterange}). In the limit of small $g_{2}$ we get
\begin{equation}
\label{fg0finiterange}
f_{g}^{(0)} (n_{0})= \frac{1}{3 \pi^{2} } \ \bigg( \frac{m g_{0} n_{0}}{\hbar^2} 
\bigg)^{3/2} \bigg[1 -  \frac{12m}{\hbar^2} g_{2} n_{0} \bigg]
\end{equation} 
This result allows us to calculate the zero-temperature number density 
$n(n_{0}, T=0)$ as the sum of the condensate density $n_{0}$ and the 
zero-temperature density contribution $f_{g}^{(0)} (n_{0})$. In particular, 
we substitute explicitly the expressions of the couplings $g_{0}$ and $g_{2}$ 
as given by Eq. (\ref{g23D}), obtaining $n(n_{0}, T=0)$ as a function of the 
three-dimensional s-wave scattering length $a_{s}$ and the effective range 
$r_{e}$:
\begin{equation}
\label{finitecondensatefraction}
n(n_{0}, T=0) = n_{0} \ \bigg[  1 + \frac{8}{3 \ \sqrt[]{\pi}} 
( n_{0} a_{s}^{3} )^{1/2} - 64 \ \sqrt[]{\pi} \frac{r_{e}}{a_{s}} 
( n_{0} a_{s}^{3} )^{3/2} \bigg]
\end{equation}
which differs from the result of reference \cite{andersen2004} by a factor 
$2$ in the finite-range correction. In Fig. \ref{fig:2} we compare our 
finite-range condensate fraction $n_{0}/n$ (black dot-dashed line) obtained 
from the numerical solution of Eq. (\ref{finitecondensatefraction}) with the 
zero-range result (blue solid line) obtained putting $r_{e} = 0$ and the 
classical result of Bogoliubov (red dashed line) \cite{bogoliubov1947}.

Let us also calculate the finite-temperature contribution $f_{g}^{(T)} (n_{0})$ 
in $D=3$, given by the integration of Eq. (\ref{fgtcondfracfiniterange}) 
in the low-temperature regime
\begin{equation}
\label{fgtfiniterangelowtemperature}
f_{g}^{(T)} (n_{0}) = \frac{(k_{B} T)^{2}}{12 (n_{0} g_{0})^{1/2}} 
\bigg( \frac{m}{\hbar^2} \bigg)^{3/2} \bigg( 1 - \frac{\pi^{2} (k_{B} T)^{2} 
\lambda(g_{2},n_{0})}{20 (n_{0} g_{0})^{2}} \bigg) + o(k_{B}T)^{5}
\end{equation}
The finite-temperature number density $n(n_{0}, T)$, according to 
Eq. (\ref{number density}), is given by the sum of the condensate density 
$n_{0}$ and the Gaussian density contributions of Eqs. (\ref{fg0finiterange}) 
and (\ref{fgtfiniterangelowtemperature})
\begin{equation}
\label{densityexpansionfiniterange3d}
n \left( n_{0},T \right) = n_{0} + \frac{1}{3 \pi^{2} } \ 
\bigg( \frac{m g_{0} n_{0}}{\hbar^2} \bigg)^{3/2} \bigg[1 -  
\frac{12m}{\hbar^2} g_{2} n_{0} \bigg] + \frac{(k_{B} T)^{2}}
{12 (n_{0} g_{0})^{1/2}} \bigg( \frac{m}{\hbar^2} \bigg)^{3/2}
\end{equation}
We also rewrite the general form of $n(n_{0}, T)$ in terms of the 
three-dimensional gas parameter $n a_{s}^{3}$ and the effective range $r_{e}$, 
employing the explicit form of the couplings $g_{0}$ and $g_{2}$ given 
by Eq. (\ref{g23D}), namely
\begin{equation}
n \left( n_{0},T \right) = n_{0} + \frac{8 (n_{0}a_{s})^{3/2}}{3 \sqrt{\pi}} 
\frac{1-8\pi^{2}n_{0}a_{s}^2r_{e}}{(1+8\pi^{2}n_{0}a_{s}^2r_{e})^2} + 
\frac{(k_{B}T)^2}{24 \sqrt{\pi}(n_{0}a_{s})^{1/2}} \bigg(\frac{m}{\hbar^2} 
\bigg)^{2}
\end{equation}
This equation can be used to express the condensate fraction $n_{0}/n$ 
explicitly in the very weakly-interacting regime in which $a_{s},r_{e} 
\rightarrow 0$, where one can approximate in the second and third subleading 
terms the condensate density $n_{0}$ with the density $n$, since the 
phenomenon of quantum depletion is absent in the noninteracting 
zero-temperature limit and these terms are finer corrections with respect 
to the first. We obtain
\beqa \nonumber
\frac{n_{0}}{n} =& 1 - \frac{8}{3 \ \sqrt[]{\pi}} ( n a_{s}^{3} )^{1/2} 
\bigg[1 - 24 \pi( n a_{s}^{3} ) \, \frac{r_{e}}{a_{s}} + 
\frac{\pi^2}{16 \zeta(3/2)^{4/3}( n a_{s}^{3} )^{2/3}}
\cdot \\ &\bigg(\frac{T}{T_{BEC}}\bigg)^2 \bigg( 1 - 
\frac{\pi^{2}}{80\zeta(3/2)^{4/3}( 
n a_{s}^{3} )^{2/3}}\bigg(\frac{T}{T_{BEC}}\bigg)^2 
\big( 1 + 8 \pi \frac{r_{e}}{a_{s}}  ( n a_{s}^3)  \big) \bigg) \bigg]
\eeqa
{where we have rescaled the temperature in terms of 
$T_{BEC}=(2\pi\hbar^{2}n^{2/3})/(mk_{B}\zeta(3/2)^{2/3})$ for 
noninteracting bosons, and $\zeta(x)$ is the Riemann zeta function.}
We emphasize that, at $T=0$ and for a zero-range interaction for which 
$r_{e} = 0$, the Bogoliubov result for the condensate quantum depletion 
is reproduced \cite{bogoliubov1947}.
Finally, we calculate the normal density $n_{n}$ of Eq. 
(\ref{normaldensityfiniterange}) substituting $D=3$ and considering 
the low-temperature regime, in which we get
\begin{equation}
\label{normaldensityfiniterange3d}
n_{n} = \frac{2 \pi^{2}}{45} \bigg(\frac{m}{\hbar^{2}}\bigg)^{3/2} 
\frac{(k_{B}T)^{4}}{(n_{0}g_{0})^{5/2}}  \bigg(1 - \frac{5 \pi^{2} 
\lambda(g_{2},n_{0}) (k_{B}T)^{2}}{2 (n_{0}g_{0})^{2}}    \bigg) + o(k_{B}T)^{7}
\end{equation}
The superfluid density $n_{s}$ for a system of bosons interacting with 
the finite-range interaction is obtained substituting 
Eqs. (\ref{finitecondensatefraction}), (\ref{fgtfiniterangelowtemperature}) 
and (\ref{normaldensityfiniterange3d}) in Eq. (\ref{superfluid density}), namely
\beqa
\label{superfluiddensityexpansion3d}
n_{s} = &n_{0} + \frac{1}{3 \pi^{2}} \ \bigg( \frac{m g_{0} n_{0}}{\hbar^2} 
\bigg)^{3/2} + \frac{(k_{B} T)^{2}}{12 (n_{0} g_{0})^{1/2}} 
\bigg( \frac{m}{\hbar^2} \bigg)^{3/2} - \nonumber 
\\ &
\frac{(8 \lambda(n_{0})+3) \pi^{2}}{360} 
\bigg(\frac{m}{\hbar^{2}}\bigg)^{3/2}  \frac{(k_{B}T)^{4}}{(n_{0}g_{0})^{5/2}}
\eeqa
As for the condensate fraction, we obtain an explicit expression of 
the superfluid fraction $n_{s}/n$ in the very weakly-interacting limit 
in which one can approximate the condensate density $n_{0}$ with the number 
density $n$ in the subleading terms, rewriting the previous equation as
\beqa \nonumber
\frac{n_{s}}{n} =& 1 -  \frac{\pi^{7/4}}{45 ( n a_{s}^{3} )^{5/6}} 
\bigg(\frac{T}{T_{BEC}}\bigg)^{4} \bigg[ 1 - \frac{5\pi^{2}}{8 \, 
\zeta(3/2)^{4/3} ( n a_{s}^{3} )^{2/3}} \bigg(\frac{T}{T_{BEC}}\bigg)^{2} 
\cdot \\ & \bigg( 1 + 8 \pi \frac{r_{e}}{a_{s}}  ( n a_{s}^3)  \bigg)  \bigg]
\label{labella}
\eeqa
where we substitute also the expressions for $g_{0}$ and $g_{2}$ of Eq. (\ref{g23D}).
Notice that, while the superfluid fraction $n_{s}/n=1$ at zero temperature, 
the condensate fraction is $n_{0}/n < 1$ due to the quantum depletion. 
{Obviously, Eq. (\ref{labella}) is reliable only in the deep 
$T/T_{BEC}\ll 1$ regime but, more generally, one must consider 
the full solution of Eq. (\ref{normaldensityfiniterange}).} 

\begin{figure}[hbtp]
\centering
\includegraphics[scale=0.7]{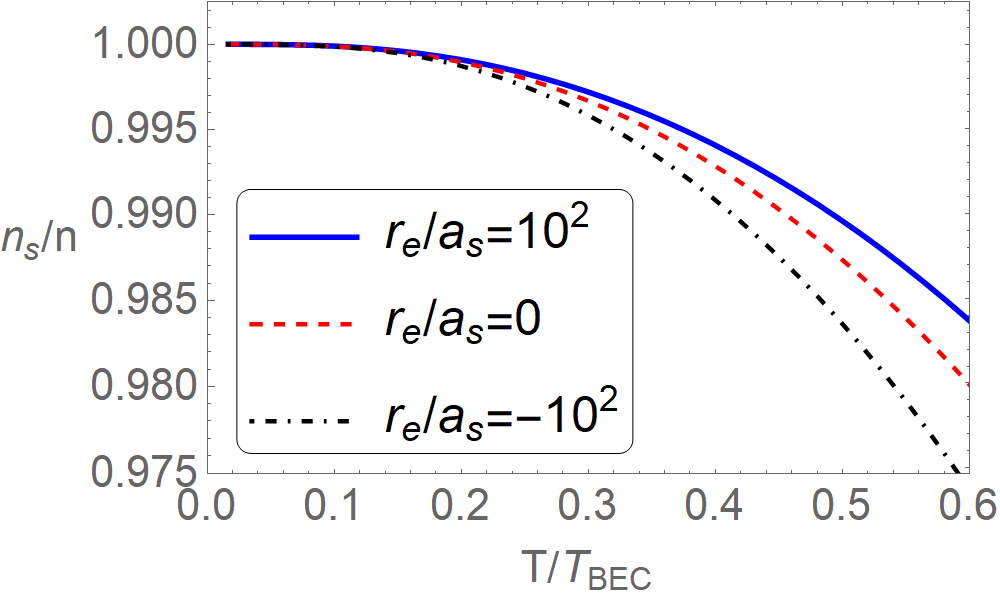}
\caption{{Superfluid fraction $n_{s}/n$ in $D=3$ for the 
gas parameter value 
$n a_{s}^{3}=10^{-4}$, obtained as the numerical integral of 
Eq. (\ref{normaldensityfiniterange}). The zero-range interaction result 
is reported as the red dashed line, while the finite-range corrections 
are reported as the solid blue line (for $r_{e}/a_{s}=10^{2}$) and 
the black dot-dashed line (for $r_{e}/a_{s}=10^{-2}$).}}
\label{fig:3}
\end{figure}

{In Fig. \ref{fig:3} we report the superfluid fraction $n_s/n$ 
as a function of the scaled temperature $T/T_{BEC}$ for three 
values of the ratio $r_e/a_s$ at fixed gas parameter $na_s^3$ 
by numerically solving Eq. (\ref{normaldensityfiniterange}). 
Fig. \ref{fig:3} shows that a positive $r_e/a_s$ slightly enhances the 
superfluid fraction while the opposite occurs for negative values.} 

\subsection{$D=2$}

\begin{figure}[ht!]
\centering
\includegraphics[scale=0.60]{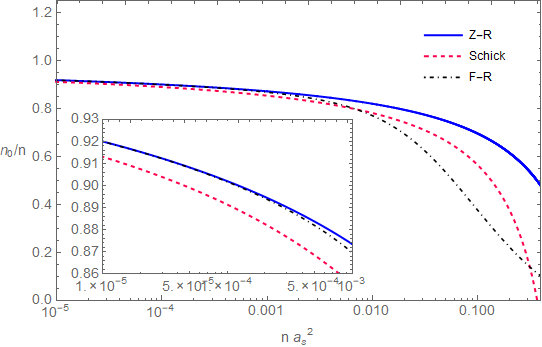}
\caption{Condensate fraction $n_{0}/n$ for bosons in $D=2$ at $T = 0$, 
reported as a function of the gas parameter $n a_{s}^{2}$. The blue solid line 
is the condensate fraction for bosons with zero-range interaction (Z-R), 
obtained from the numerical solution of Eq. 
(\ref{densityexpansionfiniterange2d}) with $R=0$. The black dot-dashed line 
is the condensate fraction for bosons with finite-range interaction (F-R) 
given by Eq. (\ref{densityexpansionfiniterange2d}) with a characteristic range 
$R$ value given by $R = 2 a_{s}$ and a ultraviolet cutoff 
$\kappa = {2 \pi }/{a_{s}}$. {The red dashed line 
is the analytical formula of Schick \cite{schick1971}, i.e. 
Eq. (\ref{lucabrutto}) with $R=0$. The inset highlights the small 
differences between our theoretical scheme and the Schick one 
in the weakly-interacting regime.}}
\label{fig:4}
\end{figure}

Here we formally follow the same path we have introduced in the 
three-dimensional case, obtaining the number density $n$ at zero-temperature 
and employing it to calculate the superfluid density $n_{s}$.
In the two-dimensional case we calculate the regularized zero-temperature 
density contribution $f_{g}^{(0)} (n_{0})$ from Eq. (\ref{fcondfracfiniterange}) 
where $D=2$ is substituted. Notice that $f_{g}^{(0)} (n_{0})$ is obtained 
as a sum of the two terms inside the square bracket of 
(\ref{fcondfracfiniterange}): while the first term is finite, the 
divergence of the second term is healed by performing a Taylor 
expansion around $\epsilon = 0$ and deleting the $o(\epsilon^{-1})$ 
divergence \cite{kleinert2001,zeidler2009}. The regularized 
zero-temperature density contribution reads
\begin{equation}
f_{g}^{(0)} (n_{0}) = \frac{m g_{0} n_{0}}{4 \pi \hbar^{2}} 
\frac{1}{\lambda(g_{2},n_{0})^{3/2}} \bigg[ \tilde{\lambda}(g_{2},n_{0}) 
+ \frac{2m}{\hbar^2} g_{2} n_{0} \ln \bigg( \frac{2 \epsilon_{0}}{ g_{0} n_{0} }
\lambda(g_{2},n_{0}) \bigg)   \bigg]
\end{equation}
where we identify the ultraviolet energy scale $\epsilon_{0}$  as
\begin{equation}
\epsilon_{0} = \frac{4 \pi \hbar^{2} \kappa^{2}  }{m e^{\gamma} }
\end{equation}
If the finite-range interaction strength $g_{2}$, as we suppose, 
constitutes a small correction of the zero-range term $g_{0}$, 
one can expand the previous equation for small values of the 
adimensional parameter $ \frac{2m}{\hbar^2} g_{2} n_{0}$, thus
\begin{equation}
f_{g}^{(0)} (n_{0}) = \frac{1}{4 \pi} \ \frac{m g_{0} n_{0}}{\hbar^{2}} + 
\frac{1}{2 \pi} \ \bigg(\frac{m }{\hbar^{2}}\bigg)^{2} g_{0} g_{2}  n_{0}^{2} 
\ \bigg[ \ln \bigg( \frac{2 \epsilon_{0}}{g_{0} n_{0} }  \bigg) -2  \bigg]
\end{equation}
In this limit, the zero-temperature density $n \left( n_{0},T=0 \right)$ reads
\begin{equation}
\label{densityexpansionfiniterange2d}
n \left(n_{0},T=0 \right) = n_{0} + \frac{1}{4 \pi} \ \frac{m g_{0} n_{0}}{\hbar^{2}} + \frac{1}{2 \pi} \ \bigg(\frac{m }{\hbar^{2}}\bigg)^{2} g_{0} g_{2}  n_{0}^{2} \ \bigg[ \ln \bigg( \frac{2 \epsilon_{0}}{g_{0} n_{0} }  \bigg) -2  \bigg]
\end{equation}
In analogy with the three-dimensional case, we derive an explicit - 
but approximated - formula for the zero-temperature condensate fraction 
$n_{0}/n$ in $D=2$ considering the very weakly-interacting regime in 
which $g_{0},g_{2} \rightarrow 0$. As before, in this regime we can 
approximate $n_{0}^2 \approx n_{0} n$ in the third term of 
Eq. (\ref{densityexpansionfiniterange2d}) and $n_{0} \approx n$ inside 
the logarithm, since it is a subleading term with respect to the first one. 
Let us come back to $\varepsilon_{0}$, by chosing the cutoff as 
$\kappa = 2 \pi /a_{s}$. Then, by taking  $g_{0}$ and $g_{2}$ as 
in Eq. (\ref{g0g22D}), we obtain the approximated condensate fraction 
$n_{0}/n$ as a function of the two-dimensional gas parameter $n a_{s}^2$ and 
the characteristic range $R$, namely
\begin{equation}
\frac{n_{0}}{n} = 1 - \frac{1}{|\ln(n a_{s}^{2})|} - 
\frac{2 \pi (R/a_{s})^{2}}{|\ln(n a_{s}^{2})|^{2}} \ \ln \bigg(  
\frac{ 8 \pi^2 |\ln(n a_{s}^{2})|}{ n a_{s}^2 e^{\gamma + 2}}  \bigg)
\label{lucabrutto}
\end{equation}
Notice that the zero-range interaction result by Schick for the condensate 
fraction $n_{0}/n$ \cite{schick1971} is easily reproduced by setting $R = 0$. 
Working outside the very weakly-interacting limit, one can also obtain the 
zero temperature condensate fraction from the numerical solution of
 Eq. (\ref{densityexpansionfiniterange2d}). We report it as the black 
dot-dashed line in Fig. \ref{fig:4}, in comparison with our 
zero-range result (blue solid line) and the result by Schick (red dashed line), 
which is reproduced in the weakly-interacting regime in which $n a_{s}^3 \ll 1$.

\noindent 
Finally, following the three-dimensional case, we may want to calculate 
also the finite-temperature density contribution $f_{g}^{(T)} (n_{0})$. 
However, substituting $D=2$ in Eq. (\ref{fgtcondfracfiniterange}) we find that
\begin{equation}
f_{g}^{(T)} (n_{0}) = \frac{m k_{B} T}{2 \pi \hbar^{2}} \int_{0}^{+\infty} dx 
\ \frac{1}{e^{x} - 1} 
\end{equation}
is infrared divergent, therefore cannot be regularized with dimensional 
regularization. This result is indeed correct and reflects the absence of 
Bose-Einstein condensation at finite-temperature in two-dimensional 
systems \cite{merminwagner1966}, as already pointed out in \cite{stoof}.

The two-dimensional normal density $n_{n} \left(n_{0},T=0 \right)$ of bosons 
with finite-range interaction is obtained from the integration of 
Eq. (\ref{normaldensityfiniterange}), in which we expand the integrand 
in the low-temperature limit, namely
\beqa
\label{normaldensityfiniterange2d} \nonumber
n_{n} \left(n_{0},T=0 \right) =& \frac{3 m}{2 \pi \hbar^{2}} 
\frac{(k_{B}T)^{3}}{(n_{0}g_{0})^{2}}  \bigg(\zeta(3) - 
\\ 
&\frac{15 \zeta(5) \lambda(g_{2},n_{0})}{(n_{0}g_{0})^{2}} (k_{B}T)^{2} 
\bigg) + o(k_{B}T)^{6}
\eeqa
Since the thermal contribution to the density $f_{g}^{(T)} (n_{0})$ is divergent 
in $D = 2$, we cannot express the superfluid density $n_{s}$ as a function of 
the condensate density $n_{0}$ at a finite temperature $T$. Therefore we 
employ the zero-temperature number density of 
Eq. (\ref{densityexpansionfiniterange2d}) to obtain the condensate density 
in the implicit form $n_{0} = n_{0} \left( n,T=0 \right)$, namely as a 
function of the density $n$. In this way we calculate the superfluid density 
substituting Eq. (\ref{normaldensityfiniterange2d}) 
into Eq. (\ref{superfluid density})
\beqa
\label{superfluiddensity2dfiniterange} \nonumber
n_{s} =& n - \frac{3 m}{2 \pi \hbar^{2}} \frac{(k_{B}T)^{3}}
{(g_{0} \, n_{0}\left(n,T=0 \right) )^{2}}  \bigg(\zeta(3) - \\ 
&\frac{15 \zeta(5) \lambda(g_{2},n_{0}\left(n,T=0 \right))}{(g_{0} \, n_{0} 
\left(n,T=0 \right))^{2}} (k_{B}T)^{2}   \bigg) 
\eeqa
which we expect to provide a reliable approximation in the low-temperature 
regime, remembering that - outside it - our approach would in any case 
be incorrect due to the Berezinski-Kosterlitz-Thouless transition 
\cite{kosterlitz1973}.

Finally, we calculate an approximated expression of the superfluid fraction 
$n_{s}/n$ in the very weakly-interacting regime where 
$g_{0},g_{2} \rightarrow 0$, in which we can approximate the condensate density 
inside Eq. (\ref{superfluiddensity2dfiniterange}) as 
$n_{0} \left(n,T=0 \right) \approx n$. In the context of this approximation, 
we also substitute in Eq. (\ref{superfluiddensity2dfiniterange}) the explicit 
form of the parameter $\lambda(g_{2},n_{0})$ of Eq. (\ref{lambda}). Moreover, 
we remember that the two-dimensional interaction strengths $g_{0}$ 
and $g_{2}$ are given by Eq. (\ref{g0g22D}), obtained in terms of the 
s-wave scattering length $a_{s}$ and the characteristic range $R$ of the 
interatomic potential with the scattering theory. The approximated superfluid 
fraction $n_{s}/n$ reads
\beqa \nonumber
\frac{n_{s}}{n} = 1 - \frac{3}{32 \pi^3} |\ln (n a_{s}^2)|^{2} \bigg(\frac{T}{T^{*}}\bigg)^{3} \bigg[ \zeta(3) \, -  \frac{15 \zeta(5)}{16 \pi^2} \bigg(\frac{T}{T^{*}}\bigg)^{2}  |\ln(n a_{s}^2)|^{2} \cdot \\
\bigg( 1 + 4 \pi \frac{R^2}{a_{s}^2} \frac{n a_{s}^2}{|\ln(n a_{s}^2)|} 
\bigg)   \bigg]
\eeqa
{where we rescale the result in terms of the temperature $T^{*}= \hbar^{2} n/(m k_{B})$ of quantum degeneracy.}
\section{Conclusions}

We have used finite-temperature one-loop functional integration to reproduce 
the density momentum equation of the two-fluid model. An analytical 
relationship between the density $n$ and the condensate density $n_{0}$ has 
been obtained at zero-temperature and in the low-temperature limit. This 
result has been used to express the low-temperature superfluid density $n_{s}$ 
as a function of $n_{0}$ and $T$ for bosons with finite-range interaction, which 
can be regarded as an explicit implementation of the Josephson relation. {
We analyze thoroughly the $D=3$ and $D=2$ case, but our approach could be applied also in
$D=1$, where we expect to reproduce the Lieb-Liniger theory except in the strong
coupling regime\cite{cappellarosalasnich1d}.}

{We expect that our theory is meaningful under the conditions of diluteness 
of the bosonic gas, for which $n a_{s}^{D} \ll 1$ and $n r_{e}^{D} \ll 1$. 
Moreover, finite-temperature results must be considered in the limit 
$k_{B}T/(g_{0} n) \ll 1$, for which the mean thermal energy is much lower 
than the gas healing length.} Our finite-range corrections to the condensate fraction 
$n_{0}/n$ and to the superfluid fraction $n_{s}/n$ can be detected in $D=3$ 
in the regime $a_{s}/r_{e} \le 1$ and in $D=2$ for $a_{s}/R \le 1$ but {\it not} 
where they are {\it much lower} than $1$. In that case, the higher order 
terms which we are neglecting in the gradient expansion of the interaction 
potential (\ref{finite-range interaction}) become relevant. Notice that, 
in $D=3$ and in $D=2$, these may represent different regimes. In fact, while 
the effective range value $r_{e}$ can be tuned by the means of a Feshbach 
resonance, the characteristice range $R$ is fixed, being 
essentially a geometric property of the real two-body interaction potential 
between the atoms. Indeed, we expect that $R$ is proportional to the Van der Waals 
radius of the atoms and it can be numerically computed, following its 
definition, using a model two-body potential $V(\vec{r})$.

An extension of this work consists in the numerical calculation of the 
thermal integrals $f_{g}^{(T)}(n_{0})$ and $n_{n} \left(n_{0},T \right)$ 
outside the zero-temperature limit in $D=2$, which we have considered for 
obtaining analytical results. In any case, we expect that our predictions fail to 
describe the superfluid fraction of the two-dimensional Bose gas at 
sufficiently high temperature, due to the occurrence of the BKT transition. 
In this case it is needed a profound rethinking of our approach, including 
explicitly in the bosonic field parametrization the contribution of vortex 
configurations of the phase field, which cause the BKT topological 
phase transion.

\section*{Acknowledgements}

The authors thank Francesco Ancilotto and Flavio Toigo for useful discussions. 
LS acknowledges for partial support the FFABR grant of 
Italian Ministry of Education, University and Research. 

{
\section*{Appendix}
In this appendix, we calculate the zero-temperature Gaussian grand potential per unit of volume $\Omega^{(0)} / L^{D}$, which is given by Eqs. (\ref{Omega0}) and (\ref{Omegag0}) in which the mean field condition Eq. (\ref{saddlepoint}) is substituted, namely
\begin{equation}
\frac{\Omega^{(0)}}{L^D}  =  -\frac{\mu_{e}^{2}}{2g} + \frac{S_{D}}{2 (2 \pi)^{D}} \int_{0}^{+\infty} dk \, k^{D-1} \ \sqrt[]{\frac{\hbar^2k^2}{2m}\bigg(\frac{\hbar^2k^2}{2m} \lambda (g_{2}, \mu_{e}/g_{0}) +2 \mu_{e} \bigg)}
\end{equation}
where $\lambda$ is defined in Eq. (\ref{lambda}). Performing dimensional regularization of the ultraviolet divergence accordingly to the superfluid density calculation, we obtain the regularized zero-temperature Gaussian grand potential $\Omega_{g}^{(0)}/ L^D$
\beqa
\label{finiterangegaussiangpregularized}\nonumber
\frac{\Omega^{(0)}}{L^D} =-\frac{\mu_{e}^{2}}{2g} - \frac{\mu_{e}^{\frac{D+2}{2}}}{2 \ \pi^{\frac{D+1}{2}}} \ \bigg( \frac{m}{\hbar^{2}} \bigg)^{D/2}   \frac{1}{\lambda (g_{2}, \mu_{e}/g_{0})^{\frac{D+1}{2}}} \bigg[ 1 + \\ \frac{\varepsilon}{2} \ln \bigg( \frac{\pi \hbar^{2} \kappa^{2} \lambda (g_{2}, \mu_{e}/g_{0})}{m \mu_{e}} \bigg) + o(\varepsilon^{2}) \bigg] \frac{\Gamma[(D - \varepsilon+1)/2] \ \Gamma[(\varepsilon -D -2)/2]}{\Gamma[(D - \varepsilon)/2]}
\eeqa
In the three-dimensional case, it is sufficient to substitute the dimension $D=3$ and put $\varepsilon = 0$ in Eq. (\ref{finiterangegaussiangpregularized}), to get the regularized zero-temperature Gaussian grand potential as
\begin{equation}
\frac{\Omega^{(0)}}{L^3} = -\frac{\mu_{e}^{2}}{2g} -\frac{\mu_{e}^{2}}{2g} \frac{8}{15 \pi^{2}} \ \bigg( \frac{m}{\hbar^{2}} \bigg)^{3/2}  \ \frac{\mu_{e}^{5/2}}{{\lambda (g_{2}, \mu_{e}/g_{0})}^{2}}
\end{equation}
which reproduces a previously known result \cite{cappellarosalasnich2017}.
In $D=2$ we expand Eq. (\ref{finiterangegaussiangpregularized}) 
and retain only $o(\varepsilon^{0})$ terms, then we identify the energy 
cutoff $\epsilon_{0}$ as
\begin{equation}
\epsilon_{0} = \frac{4 \pi \hbar^{2} \kappa^{2}  }
{m \exp(\gamma + 1 - \frac{4 \pi \hbar^{2} {\lambda(g_{2}, 
\mu_{e}/g_{0})}^{3/2}}{m g_{0}}) }
\end{equation}
to get the zero-temperature Gaussian grand potential
\begin{equation}
\label{grandpotential0Tfiniterange}
\frac{\Omega^{(0)}}{L^2} = - \frac{m \mu_{e}^{2}}{8 \pi \hbar^{2} 
{\lambda(g_{2}, \mu_{e}/g_{0})}^{3/2}} \ \bigg[ \ln \bigg( \frac{\epsilon_{0}}
{\mu_{e}}\lambda (g_{2}, \mu_{e}/g_{0}) \bigg) + \frac{1}{2} \bigg]  
\end{equation}
This equation of state corrects the one obtained 
in Ref. \cite{salasnich2017bis}: here $\lambda (g_{2}, \mu_{e}/g_{0})$ 
appears also inside the logarithm. Moreover, it is important to stress that, 
in the case of zero-range interaction, Mora and Castin \cite{moracastin} 
were able to extend Eq. (\ref{grandpotential0Tfiniterange}) by including 
a beyond-Gaussian term. This next-next-to-leading extension 
in the finite-range case is highly non trivial and it surely deserves 
a separate detailed investigation.}


\section*{References}

\begin{thebibliography}{99}

\bibitem{tilley} Tilley D R and Tilley J 1990
\textit{Superfluidity and Superconductivity}
(CRC Press)

\bibitem{pezze2018} Pezz\'e L, Smerzi A, Oberthaler M K, Schmied M and 
Treutlein P 2018 \textit{Rev. Mod. Phys.} \textbf{90} 035005

\bibitem{streltsov2017} Streltsov A, Adesso G and Plenio M 2017
\textit{Rev. Mod. Phys.} \textbf{89} 041003

\bibitem{kimble2018} Chang D E, Douglas J S, Gonzalez-Tudela A, 
Hung C L and Kimble H J 2018 \textit{Rev. Mod. Phys.} \textbf{90} 031002

\bibitem{bcs} Bardeen J, Cooper L N and Schrieffer J R 1957
\textit{Phys. Rev.} \textbf{106} 1

\bibitem{varlamov-book} Larkin A and Varlamov A 2005
\textit{Theory of fluctuations in superconductors}
(Oxford: Oxford University Press)

\bibitem{hohenberg1965} Hohenberg P C and Martin P C 1965
\textit{Annals of Physics} \textbf{281} 636-705

\bibitem{anderson1966} Anderson P W 1966
\textit{Rev. Mod. Phys.} \textbf{38} 2

\bibitem{ceperley1995} Ceperley D M 1995
\textit{Rev. Mod. Phys.} \textbf{67} 279

\bibitem{schmitt-book} Schmitt A 2014
\textit{Introduction to Superfluidity: Field-theoretical 
Approach and Applications} (Switzerland: Springer)

\bibitem{landau1941} Landau L D 1941 { \it Phys. Rev.} \textbf{60} 356

\bibitem{landau1981} Landau L D and Lifshitz E M 1981
{ \it Statistical Physics - Part 2} (Oxford: Pergamon Press)

\bibitem{leggett-book} Leggett A J 2006
\textit{Quantum Liquids: Bose Condensation and Cooper Pairing 
in Condensed-matter Systems} (Oxford: Oxford University Press)

\bibitem{london1938-1} London F 1938
\textit{Nature} \textbf{141} 643

\bibitem{london1938-2} London F 1938
\textit{Phys. Rev.} \textbf{54} 937

\bibitem{bogoliubov1947} Bogoliubov N N 1947 
\textit{J. Phys. U.S.S.R.} \textbf{11} 23

\bibitem{huang-book} Huang K 1987
\textit{Statistical Mechanics} (Wiley \& sons)

\bibitem{stringari-book} Stringari S and Pitaevskii L P 2016
\textit{Bose-Einstein Condensation and Superfluidity}
(Oxford University Press)

\bibitem{hadzibabic2009} Hadzibabic Z and Dalibard J 2011
\textit{Rivista del Nuovo Cimento} \textbf{34} 389

\bibitem{dalibard2011} Dalibard J, Gerbier F,  Juzeliunas G and \"Ohberg P 2011
\textit{Rev. Mod Phys.} \textbf{83} 1523

\bibitem{braaten2001} Braaten E,  Hammer H W and Hermans S 2001
\textit{Phys. Rev. A} \textbf{63} 063609

\bibitem{andersen2004} Andersen J O 2004
\textit{Rev. Mod. Phys.} \textbf{76} 599

\bibitem{cappellarosalasnich2017} Cappellaro A and Salasnich L 2017 
{\it Phys. Rev. A } \textbf{95} 033627

\bibitem{yukalov} Yukalov V I and Yukalova E P 2016 { \it Laser Phys.} 
\textbf{26} 045501

\bibitem{salasnich2017bis} Salasnich L 2017 {\it Phys. Rev. Lett. } 
\textbf{118} 130402

\bibitem{bean} Beane S R 2018 { \it Eur. Phys. J. D} 
\textbf{72} 55

\bibitem{schakel} Schakel A M J 2008
{\it Boulevard of Broken Symmetries} (Singapore: World Scientific)

\bibitem{fisher} Fisher M E, Barber M and Jasnow D
1973 { \it Phys. Rev. A } \textbf{8} 1111

\bibitem{taylor2006} Taylor E, Griffin A, Fukushima N and Ohashi Y, (2006),
Phys. Rev. A \textbf{74} 063626

\bibitem{nagaosa} Nagaosa N 1999 { \it Quantum Field Theory in 
Condensed Matter Physics} (Springer: Verlag)

\bibitem{boudjemaa2011} 
Boudjem\^{a}a A and Benarous M 2011 {\it Phys. Rev. A} \textbf{84} 043633

\bibitem{boudjemaa2012}
Boudjem\^{a}a A 2012 {\it Phys. Rev. A} \textbf{86} 043608

\bibitem{boudjemaa2017}
Boudjemaa A and Guebli N 2017 {\it J. Phys. A: Math Theor.} \textbf{50} 425004

\bibitem{altland} Altland A and Simons B D 2010 {\it  Condensed Matter 
Field Theory } (Cambridge: Cambridge University Press)

\bibitem{salasnichtoigo} Salasnich L and Toigo F 2016
{\it  Phys. Rep.} \textbf{640} 1-29 

\bibitem{josephson1966} Josephson B D 1966 {\it Phys. Lett.  } \textbf{21} 608

\bibitem{svistunov} Svistunov B, Babaev E, Prokof'ev N 2015 
{\it Superfluid States of Matter} (CRC Press: Boca Raton)

\bibitem{astrakharchik2009} Astrakharchik G E,  Boronat J, Casulleras J, 
Kurbakov I L, and Lozovik Yu E 2009 {\it Phys. Rev. A} \textbf{79} 051602

\bibitem{thooft1972}
't Hooft G and Veltman M 1972  {\it Nucl. Phys. B} \textbf{44} 189-213

\bibitem{leibbrandt1975} Leibbrandt G 1975 {\it Rev. Mod. Phys. } 
\textbf{47} 849

\bibitem{kleinert2001} Kleinert H, Schulte-Frohlinde V 2001 
{ \it Critical Properties of $\phi^{4}$ Theories} (Singapore: World Scientific)

\bibitem{zeidler2009} Zeidler E 2009 {\it Quantum Field Theory II: 
Quantum Electrodynamics} (Berlin: Springer)

\bibitem{schick1971} Schick M 1971 { \it Phys. Rev. A }  \textbf{3} 1067

\bibitem{merminwagner1966} Mermin N D and Wagner H
1966 { \it Phys. Rev. Lett.} \textbf{17} 1133

\bibitem{stoof}
Stoof H T C, Dickerscheid  D B M and Gubbels K 2009 {\it Ultracold 
Quantum Fields} (Dordrecht: Springer)

\bibitem{kosterlitz1973}
Kosterlitz L M and Thouless D J 1973 {\it J. Phys. C} \textbf{6} 1181

\bibitem{cappellarosalasnich1d}
Cappellaro A and Salasnich L 2017 {\it Phys. Rev. A} \textbf{96} 063610

\bibitem{moracastin}
Mora C and Castin Y 2009 {\it Phys. Rev. Lett.} \textbf{102} 180404

\end{thebibliography}
\end{document}